\documentclass[12pt]{article}
\usepackage{amsfonts,latexsym,graphicx}
\DeclareMathAlphabet{\mathpzc}{OT1}{pzc}{m}{it}
\input xy
\xyoption{all}
\usepackage{graphics}
\usepackage{amsfonts,latexsym}
\DeclareMathAlphabet{\mathpzc}{OT1}{pzc}{m}{it}
\begin{document}
\title{An algebraic origin for quark confinement}
\author{P.S. Isaac\\ Graduate School of Mathematical Sciences, \\
The
University of Tokyo, \\   3-8-1 Komaba, Meguro, 153-8914,
Tokyo, Japan \\ email: isaac@ms.u-tokyo.ac.jp 
\\~~ \\   W.P. Joyce \\ Department of Physics and Astronomy,
\\ The University of
Canterbury, \\ 
Private Bag 4800, Christchurch, New Zealand   
\\ email: william.joyce@canterbury.ac.nz \\
~~ \\ J. Links \\  Department of Mathematics, \\ 
The University of
Queensland, \\ Brisbane, 4072, Australia \\ email: jrl@maths.uq.edu.au  
}                     
%
%
%
\maketitle
\begin{abstract}
Recently, a principle for state confinement has been  
proposed in a category theoretic framework and to accomodate this result the
notion of a pre-monoidal category was developed. Here we describe an  
algebraic approach for the construction of such categories.  
We introduce a procedure called twining which breaks the quasi-bialgebra
structure of the universal enveloping algebras of semi-simple Lie
algebras and renders the category of finite-dimensional modules
pre-monoidal. The category is also symmetric, meaning that each
object of the category provides representations of the symmetric groups,
which allows for a generalised
boson-fermion statistic to be defined. Exclusion and confinement
principles for systems of indistinguishable particles 
are formulated as an invariance with respect to the symmetric
group actions. We apply the construction to several examples and in
particular show that the symmetries which can be associated to
colour, spin and flavour degrees of freedom of quarks do lead naturally to
confinement of states. 
\end{abstract}

\vfil\eject


\def\a{\alpha}
\def\b{\beta}
\def\d{\delta}
\def\e{\epsilon}
\def\ve{\varepsilon}
\def\g{\gamma}
\def\k{\kappa}
\def\l{\lambda}
\def\o{\omega}
\def\t{\theta}
\def\s{\sigma}
\def\D{\Delta}
\def\L{\Lambda}
\def\X{\bar{X}}
\def\Y{\bar{Y}}
\def\Z{\bar{Z}} 
\def\ch{\check}
\def\f{{\cal F}} 
\def\G{{\cal G}}
\def\hG{{\hat{\cal G}}}
\def\R{{\cal R}}
\def\hR{{\hat{\cal R}}}
\def\tR{{\tilde{\cal R}}}
\def\C{{\mathbf C}}
\def\P{{\bf P}}
\def\T{{\cal T}}
\def\H{{\cal H}}
\def\trho{{\tilde{\rho}}}
\def\tP{{\tilde{\Phi}}}
\def\tT{{\tilde{\cal T}}}
\def\ot{\otimes} 
\def\aa{{\mathpzc{A}}}
\def\qq{{\mathpzc{Q}}}
\def\l{\left} 
\def\r{\right}

\def\tp{\otimes}

\def\I{{\rm id}}

\def\ev{{\rm ev}}

\def\coev{{\rm coev}}

\def\rar{\rightarrow}

\def\idel{{\mathbf 1}}


\def\beq{\begin{equation}}
\def\eeq{\end{equation}}
\def\bea{\begin{eqnarray}}
\def\eea{\end{eqnarray}}
\def\ba{\begin{array}}
\def\ea{\end{array}}
\def\no{\nonumber}
\def\lt{\left}
\def\rt{\right}
\newcommand{\bq}{\begin{quote}}
\newcommand{\eq}{\end{quote}}

\newtheorem{Theorem}{Theorem}
\newtheorem{Definition}{Definition}
\newtheorem{Proposition}{Proposition}
\newtheorem{Lemma}{Lemma}
\newtheorem{Corollary}{Corollary}

\section{Introduction} 

Despite many successes, the quark model of elementary particle physics
\cite{gn,fs,huang} poses some puzzling questions, none more so than the issue of
state confinement. Quarks carry the quantum label {\em colour} which can
take one of three values; red, green or blue. Accordingly, it is
natural to associate $su(3)$ symmetry to describe invariance with
respect to transformations of colour. The introduction of
colour was to ensure that the states for particles such as protons and neutrons
possessed the correct statistics (antisymmetry of the wavefunction with
regard to quark interchange) \cite{hn}. 
Colour also plays the pivotal role of ``charge'' in the theory of 
quantum chromodynamics \cite{lm,wilczek}. 
However, 
all particles observed to date are colour neutral, such as the baryons
(including protons and neutrons) 
composed of three quarks, each with different colour index, or the mesons
comprised of combinations of quarks and antiquarks in colour-anticolour
pairs.    
This has lead to the principle of quark confinement. In
turn, it has been postulated that a ``confining force'' exists to
explain this phenomenon, although a clear picture about the nature of
such a force is lacking. It is believed that for
extremely 
high temperatures a deconfinement phase should prevail, giving rise to a
quark-gluon plasma where colour might be observed. 
Possible examples of this are the early stages of the
universe and neutron stars.   
There is also hope that this state of
matter can be observed in the laboratory \cite{qr}.  

Recently,   
an alternative approach to explain
confinement as a consequence of symmetry rather than dynamics has been
put forth \cite {joyce-s}. This  
idea that confinement follows from symmetry 
has been proposed previously \cite{huang,pol}, although from a different
standpoint where the centre (or Casimir invariants) of the $su(3)$ algebra are
employed. In \cite{joyce-s} the problem is set in a category theory context;
viz. the collection of finite-dimensional modules of $su(3)$, coupled
with a statistic which identifies states as being fermionic or
bosonic, is a category. It was identified that the only category which was
compatible with the boson-fermion statistic was a symmetric pre-monoidal
structure \cite{joyce-s,joyce-q,joyce-b,joyce-v}. 
A feature of this structure is that associativity is broken.
An indication why this is necessary can be seen from the following
argument. The three quark colour indices provide the basis vectors for 
the fundamental 3-dimensional
representation of the $su(3)$ algebra, which we denote $\{3\}$. The
anticolour indices are associated with the basis vectors of the
dual representation, denoted
$\{\overline{3}\}$. It is required that all the states of both the
fundamental module and its dual are fermionic, which was the prime
motivation for introducing colour. A contradiction seems to result from the
well known $su(3)$ decomposition 
$$ \{3\}\otimes \{3\}=\{6\}\oplus \{\overline{3}\}. $$  
That is, the fermionic antiquark states are
isomorphic to the antisymmetrised  
two-fold tensor products of fermionic states which afford the
representation $\{3\}$. Taking the
usual approach that the statistics of particles is additive, this leads
to a contradiction since two fermions should combine to form a (hard-core)
bosonic state. 
One solution to this problem is to replace the statistic with
a ${\mathbb Z}_3$ parastatistic \cite{para,tony}. The solution promoted
in \cite{joyce-s} is to maintain the boson-fermion ${\mathbb Z}_2$ 
statistic and instead 
break the associativity of the tensor products. An advantage of this
approach is that it then allows for a natural characterisation of confinement.

There exists a rich mathematical theory which provides a mechanism to 
construct non-associative tensor product 
structures. This is provided
by the theory of quasi-bialgebras \cite{drinfeld1,drinfeld2}. 
The quasi-bialgebras possess a generically
non-associative tensor algebra (in some of the literature 
this is referred to as a
non co-associative co-algebra structure). Through representations of
quasi-bialgebras, non-associative tensor products of vector spaces are
obtained, which form a monoidal category. 
There exists also the procedure of {\em twisting} which
allows many non co-associative co-algebra structures to be assigned to a
single algebra. The categories of modules are equivalent when
the co-algebra structures are related by twisting. In cases for which
the quasi-bialgebra is quasi-triangular this category becomes  
{\it braided} \cite{js}. Braiding may be thought of as a means for
describing the statistics of
the states, a generalisation of the familiar notion of the boson-fermion
statistic. 
However, as the category of finite-dimensional modules of any 
quasi-bialgebra
is always, by construction, monoidal, this means that the
pentagon axiom \cite{mc} is always obeyed. The results of \cite{joyce-s}
showed that
the only possibility for a boson-fermion statistic for $su(3)$ colour is
for the modules to form a strictly pre-monoidal category, corresponding to a
deformation of the pentagon axiom. 

The aim of this paper is to demonstrate that there indeed exists an
algebraic procedure, to be called {\it twining}, which breaks the 
quasi-bialgebra structure and permits the
construction of pre-monoidal categories of modules. Although the
procedure may not always extend to the construction of braided, pre-monoidal
categories of modules,
we will show that it is always possible for the cases when the modules belong
to 
the universal enveloping algebras of semi-simple Lie algebras. 
The application of
twining utilises the Casimir invariants of the Lie algebra, which brings
together the ideas of both \cite{joyce-s} and \cite{huang,pol} 
to formulate state
confinement as a consequence of symmetry and statistics. This principle
is formulated as an invariance with respect to the symmetric group
action defined on the category of modules. As examples we
study the case of $su(2)$, for which the construction yields the
spin statistics theorem; i.e. states of even-dimensional $su(2)$ modules
are fermionic while those of odd-dimensional modules are bosonic. Extending
this analysis to $su(3)$ leads to a reproduction of the results of
\cite{joyce-s}, from which colour confinement is deduced. For the case of the
semi-simple Lie algebra $su(2)\oplus su(3)$, which provide the symmetry
algebra for light quark spin and flavour, the results are consistent
with the classification of hadrons due to Gell-Mann and Ne'eman, 
known as the 
{\em Eightfold Way} \cite{gn}. 

\section{Pre-monoidal, monoidal and braided categories}\label{cats}  

This section gives a brief summary of the category theoretic concepts which
will be referred to in this paper.
Let us first begin with the definition of a pre-monoidal category, taken
from \cite{joyce-q}. (See also \cite{yanofsky}).
\begin{Definition}\label{pmc}
A pre-monoidal category is a triple $(\C,\otimes,\aa)$ where $\C$ is a
category, $\otimes:\C\times \C\rightarrow \C$ is a bifunctor and
$\aa:\otimes({\rm id}\times\otimes )\rightarrow\otimes( \otimes\times{\rm id})$ is a
natural isomorphism for associativity. 
\end{Definition} 
We refer to ${a}_{U,V,W}$ as the component isomorphisms of $\aa$
for all objects $U,V,W\in \C$.
Therefore $a_{U,V,W}$ is the morphism $a_{U,V,W}: U\tp (V\tp W)\rar (U\tp V)\tp W $.
We note that at this stage there are no conditions imposed on
$\aa$, but it is useful to define the natural isomorphism 
$\qq:\otimes(\otimes\times
\otimes)\rightarrow \otimes(\otimes\times\otimes)$ through the following
diagram 
$$
\xymatrix{
(U\tp V)\tp(W\tp Z)  \ar[dd]_{a_{(U\tp V),W,Z}}                           &
                                                                        & 
(U\tp V)\tp(W\tp Z) \ar[ll]_{q_{U,V,W,Z}}                               \\ 
&&\\
((U\tp V)\tp W)\tp Z                                                    &
                                                                        &
U\tp (V\tp (W\tp Z)) \ar[uu]_{a_{U,V,(W\tp Z)}} \ar[dd]^{\I\tp a_{V,W,Z}} \\
&&\\
(U\tp(V\tp W))\tp Z \ar[uu]^{a_{U,V,W}\tp \I}                           &
                                                                        &
U\tp ((V\tp W)\tp Z) \ar[ll]^{a_{U,(V\tp W),Z}}    
}
$$

In the above diagram we have written $\qq$ in terms of component 
isomorphisms defined by 
$$\qq\left((U\otimes V)\otimes(W\otimes Z)\right) 
=q^{}_{U,V,W,Z}\left((U\otimes V)\otimes(W\otimes
Z)\right)$$ 
so we may write
$$q^{}_{U,V,W,Z}=a^{-1}_{(U\otimes V),W,Z}(a^{}_{U,V,W}\tp \I)
a^{}_{U,(V\otimes W),Z} 
(\I \tp a^{}_{V,W,Z})a^{-1}_{U,V,(W\otimes Z)}. $$ 
The above diagram can be viewed as a generalisation of the pentagon diagram
in the definition of monoidal categories. 
In order to give a mathematical interpretation of the significance of
$\qq$ it is useful to distinguish the coupling of the objects of the
category through the brackets $[~],~\{~\}$. Adopting this notation it can be
shown that 
$$\qq([U\tp V]\tp \{W\tp Z\})=(\{U\tp V\}\tp[ W\tp Z])$$ 
so the functor $\qq$ provides a morphism for ``temporal coupling'' where
the coupling of $U$ and $V$ {\em before} coupling $W$ and $Z$ is
distinguished from the coupling in the reverse order. A representation for
such a temporal coupling scheme in terms of binary trees is given in
\cite{joyce-q}.

Of particular relevance
to this paper are pre-monoidal categories with the following
property.

\begin{Definition}\label{upmc}
A pre-monoidal category is called unital if 
there is an identity object ${\mathbf 1}\in \C$ and natural
isomorphisms $\rho_U: U\otimes {\mathbf 1}\rightarrow U$ and $\lambda_U:
{\mathbf 1}\otimes U\rightarrow U$ such that the diagrams
\begin{itemize}
\item[(i)]
$$
\xymatrix{
(U\tp {\mathbf 1})\tp V  \ar[ddrr]_{\rho_U\tp\I}                             &
                                                                             &
U\tp({\mathbf 1}\tp V) \ar[ll]_{a_{U,{\mathbf 1},V}}\ar[dd]^{\I\tp\lambda_V} \\
&&\\
                                                                             &
                                                                             &
U\tp V
}
$$
\item[(ii)]
$$
\xymatrix{
({\mathbf 1}\tp U)\tp V  \ar[ddrr]_{\lambda_U\tp\I}                             &
                                                                             &
{\mathbf 1}\tp(U\tp V) \ar[ll]_{a_{{\mathbf 1},U,V}}\ar[dd]^{\lambda_{U\tp V}} \\
&&\\
                                                                             &
                                                                             &
U\tp V
}
$$
\item[(iii)]
$$
\xymatrix{
(U\tp V)\tp {\mathbf 1}  \ar[ddrr]_{\rho_{U\tp V}}                             &
                                                                             &
U\tp(V\tp {\mathbf 1}) \ar[ll]_{a_{U,V,{\mathbf 1}}}\ar[dd]^{\I\tp\rho_V} \\
&&\\
                                                                             &
                                                                             &
U\tp V
}
$$
\end{itemize}
commute for all $U,V$.  
\end{Definition}
To complete the picture it should be clear that a unital, 
pre-monoidal category for which 
$$
\qq=\I\tp\I\tp\I\tp\I
$$ 
is a monoidal category.
Monoidal categories are defined in \cite{mc}. Discussions relevant
to the present work can be found in \cite{cp,majid}. 

Another important notion to consider is that of rigidity. Rigidity is
related to the inclusion of dual objects in the category in a consistent
way. An object $U^*$ is said to be dual to an object $U$ in a unital, 
pre-monoidal
category $\C$ if there exist morphisms $\ev_U:U^*\tp U\rar \idel$ and 
$\coev_U:\idel\rar U\tp U^*$ such that the following diagrams commute.

$$
\xymatrix{
\idel\tp U  \ar[rr]^{\coev_U\tp\I}                          &&
(U\tp U^*)\tp U                                           \\
U\ar[u]^{\lambda^{-1}_U}                                  &&
                                                          \\
U\tp\idel \ar[u]^{\rho_U}                                 &&
U\tp(U^*\tp U)\ar[ll]^{\I\tp \ev_U} \ar[uu]_{a_{U,U^*,U}}               
}
$$ 

$$
\xymatrix{
U^*\tp\idel\ar[d]_{\rho_{U^*}} \ar[rr]^{\I\tp \coev_U} &&
U^*\tp(U\tp U^*)  \ar[dd]^{a_{U^*,U,U^*}}           \\
U^*                                                 &&
                                                    \\
\idel\tp U^* \ar[u]^{\lambda_{U^*}}                 &&
(U^*\tp U)\tp U^*  \ar[ll]^{\ev_U\tp\I}
}
$$
Assuming the existence of such morphisms, for any morphism 
$\psi:U\rar V$, we are able to define its dual $\psi^*:V^*\rar U^*$
by the commutative diagram
$$
\xymatrix{
V^*\tp \idel \ar[rr]^{\I\tp \coev_U}             &&
V^*\tp(U\tp U^*) \ar[dd]^{\I\tp\psi\tp\I}     \\ 
V^* \ar[u]^{\rho^{-1}_{V^*}} \ar[dd]_{\psi^*} &&
                                              \\
                                              &&
V^*\tp(V\tp U^*) \ar[dd]^{a_{V^*,V,U^*}}      \\
U^*                                           &&
                                              \\
\idel\tp U^* \ar[u]^{\lambda_{U^*}}           &&
(V^*\tp V)\tp U^* \ar[ll]^{\ev_V\tp\I}        
}
$$
The contravariant functor which maps all objects $U$ into their
dual $U^*$ and morphisms $\psi$ into $\psi^*$ by the construction
of the above diagram is referred to as the dual object functor.

We then say that a unital, pre-monoidal category 
is {\em rigid} if every object has a dual
object and if the dual object functor is an (anti-)equivalence of
categories. The above discussion on rigidity and duality can also be found 
in \cite{cp} but we have included it here for the reader's convenience.

An important class of unital pre-monoidal categories is where 
the tensor product is
commutative up to isomorphism. This leads to the definition of {\em braided},
(sometimes in the literature referred to as {\em quasi-tensor},)
unital, pre-monoidal categories which we provide below. Coherence for
these categories was determined in \cite{joyce-b}.

\begin{Definition}\label{bmc}
A unital pre-monoidal category $\C$ is said to be braided if it is equipped with
a natural commutativity isomorphism $\s_{U,V}:U\tp V \rar V\tp
U$ for all objects $U,V\in \C$ such that the following diagrams
commute:
\begin{itemize}
\item[(i)] 
$$
\xymatrix{
                                                                & 
U\tp (V\tp W) \ar[rr]^{\s_{U,(V\tp W)}}                           &&
(V\tp W)\tp U \ar[dr]^{a^{-1}_{V,W,U}}                         &               
                                                                \\
(U\tp V)\tp W \ar[ur]^{a^{-1}_{U,V,W}} \ar[dr]_{\s_{U,V}\tp\I} &&&&
V\tp (W\tp U)                                                   \\
                                                                &
(V\tp U)\tp W \ar[rr]_{a^{-1}_{V,U,W}}                         &&
V\tp (U\tp W) \ar[ur]_{\I\tp\s_{U,W}}                           &
}
$$ 
\item[(ii)]
$$
\xymatrix{
                                                           & 
(U\tp V)\tp W \ar[rr]^{\s_{(U\tp V),W}}                      &&
W\tp (U\tp V) \ar[dr]^{a_{W,U,V}}                         &               
                                                           \\
U\tp (V\tp W) \ar[ur]^{a_{U,V,W}} \ar[dr]_{\I\tp\s_{V,W}} &&&&
(W\tp U)\tp V                                              \\
                                                           &
U\tp (W\tp V) \ar[rr]_{a_{U,W,V}}                         &&
(U\tp W)\tp V \ar[ur]_{\s_{U,W}\tp\I}                      &
}
$$
\item[(iii)]
$$
\xymatrix{
(U\tp V)\tp (W\tp Z) \ar[d]_{\s_{(U\tp V),(W\tp Z)}} \ar[rr]^{q_{U,V,W,Z}} &&
(U\tp V)\tp (W\tp Z) \ar[d]^{\s_{(U\tp V),(W\tp Z)}} \\
(W\tp Z)\tp (U\tp V) && (W\tp Z)\tp (U\tp V) \ar[ll]_{q_{W,Z,U,V}}
}
$$
\end{itemize}
\end{Definition}
Note that the triangle diagrams with $(i)$ and $(ii)$ above imply that
$\s_{\idel ,U}=\rho^{-1}_{U}\lambda^{}_{U}$ and
$\s_{U,1}=\lambda^{-1}_{U}\rho^{}_{U}$. Using the invertibility of $\rho$
and $\lambda$, we obtain the result $\s^{-1}_{1,U}=\s^{}_{U,1}$. In
general, if $\s_{U,V}\circ\s_{V,U}=\I_{V\tp U}$ for all objects $U,V\in
\C$, $\C$ is said to be {\em symmetric} (occasionally in the literature
referred to as {\em tensor}). In this case the diagrams $(i)$ and $(ii)$
are equivalent.

\section{Quasi-bialgebras, quasi-Hopf algebras and quasi-triangularity} 

A systematic way to obtain monoidal categories is through
representations of quasi-bialgebras, defined below. 
Throughout we take the underlying field to be the complex numbers
$\mathbb C$. Let $A$ be an associative algebra with multiplication 
$m: A\tp A\rightarrow A$ denoted $m(a\otimes b)=ab,~
\forall a,b\in A$. Since $A$ is associative we have
$m\cdot(m\otimes \I)=m\cdot (\I\otimes m)$. 
The following is due to Drinfel'd \cite{drinfeld1,drinfeld2}.  
\begin{Definition}\label{quasi-bi}: 
A complex quasi-bialgebra $(A,\D,\e,\Phi)$ is an associative algebra $A$
over $\mathbb C$  with unit element $I$, 
equipped with algebra homomorphisms $\e: 
A\rightarrow {\mathbb C}$ (co-unit), $\D: A\rightarrow A\otimes A$ (co-product),
and an invertible element $\Phi\in A\otimes A\otimes A$ 
(co-associator) satisfying 
\bea
&& (\I\otimes\D)\D(a)=\Phi^{-1}(\D\otimes \I)\D(a)\Phi,~~
      \forall a\in A,\label{quasi-bi1}\\
&&(\D\otimes \I\otimes \I)\Phi \cdot (\I\otimes \I\otimes\D)\Phi
=(\Phi\otimes I)\cdot(\I\otimes\D\otimes \I)\Phi\cdot (I\otimes
                  \Phi),\label{quasi-bi2} \label{pent}\\
&&m\cdot(\e\otimes \I)\D=\I=m\cdot(\I\otimes\e)\D,\label{quasi-bi3}\\
&&m\cdot(m\tp \I)\cdot
(\I\otimes\e\otimes \I)\Phi=I. \label{quasi-bi4} \eea  
If, moreover, there exists  an 
algebra anti-homomorphism $S: A\rightarrow A$ (anti-pode) and 
canonical elements $\a,~\b\in A$, satisfying 
\bea
&& m\cdot (I\otimes\a)(S\otimes \I)\D(a)=\e(a)\a,~~~\forall
    a\in A,\label{quasi-hopf1}\\
&& m\cdot (I \otimes\b)(\I\otimes S)\D(a)=\e(a)\b,~~~\forall a\in A,
     \label{quasi-hopf2}\\
&& m\cdot (m\otimes \I)\cdot (I \otimes\b\otimes\a)(\I\otimes S\otimes
     \I)\Phi^{-1}=I  ,\label{quasi-hopf3}\\
&& m\cdot(m\otimes \I)\cdot (I\otimes\a\otimes
     \b)(S\otimes \I\otimes S)\Phi=I \label{quasi-hopf4}
\eea
then $(A,\D,\e,\Phi,S,\alpha,\beta)$ 
is said to be a quasi-Hopf algebra. 
If there
exists an invertible element $\R\in A\otimes A$ (universal $R$-matrix) 
such that
\bea
&&\R\D(a)=\D^T(a)\R,~~~~\forall a\in A,\label{dr=rd}\\
&&(\D\otimes
\I)\R=\Phi^{-1}_{231}\R_{13}\Phi_{132}\R_{23}\Phi^{-1}_{123},
   \label{d1r}\\
   &&(\I\otimes \D)\R=\Phi_{312}\R_{13}\Phi^{-1}_{213}\R_{12}\Phi_{123}
      \label{1dr}
      \eea
then $(A,\D,\e,\Phi,R)$ is called a quasi-triangular quasi-bialgebra. 
\end{Definition}
For all elements $a,b\in A$, the antipode satisfies
\beq
S(ab)=S(b)S(a).  
\eeq
The equations (\ref{quasi-bi2}), (\ref{quasi-bi3}) and (\ref{quasi-bi4}) imply
that $\Phi$ also obeys
\bea 
&&m\cdot (m \tp \I)\cdot (\e\otimes \I\otimes \I)\Phi\no \\
&&~~=m\cdot (m \tp \I)\cdot(\I\otimes \I\otimes\e) 
\Phi \no \\
&&~~=I.\label{e(phi)=1} \eea 
Applying $\e$ to the axioms (\ref{quasi-hopf3}, \ref{quasi-hopf4}) 
we obtain, in view
of (\ref{quasi-bi4}), $\e(\a)\e(\b)=1$. 
By applying $\e$ to
(\ref{quasi-hopf1}), we have $\e(S(a))=\e(a),~\forall a\in A$.
The fact that $\e$ is a homomorphism means that $\e$ defines a
one-dimensional representation for $A$. 

Above, $\D^T=T\cdot\D$ with $T$ being the twist map
which is defined by
\beq
T(a\otimes b)=b\otimes a;
\eeq
and $\Phi_{132}$  etc. are derived from $\Phi\equiv\Phi_{123}$
through use of $T$
\bea
&&\Phi_{132}=(\I\otimes T)\Phi_{123},\no\\
&&\Phi_{312}=(T\otimes \I)\Phi_{132}=(T\otimes \I)
   (\I\otimes T)\Phi_{123},\no\\
   &&\Phi^{-1}_{231}=(\I\otimes T)\Phi^{-1}_{213}=(\I\otimes T)
      (T\otimes \I)\Phi^{-1}_{123},\no
      \eea
and so on, and similarly for $R_{13}$ etc.  
We mention that our convention for the definitions of $\Phi_{132}$
{\it etc} follow \cite{cp,majid}. The convention is not uniformly adopted
throughout the literature (e.g. \cite{ac}).
It is easily shown that the properties (\ref{dr=rd})-(\ref{1dr}) imply
\beq \R_{12}\Phi^{-1}_{231}\R_{13}\Phi_{132}\R_{23}\Phi^{-1}_{123}
      =\Phi^{-1}_{321}\R_{23}\Phi_{312}\R_{13}\Phi^{-1}_{213}\R_{12},
        \label{qybe}
        \eeq
which is referred to as the  quasi-Yang-Baxter equation.

The category of quasi-bialgebras
is invariant under a form of gauge transformation known as twisting. 
Let $(A,\D,\e,\Phi)$
be a quasi-bialgebra
and let $F\in A\otimes A$
be an invertible element satisfying the co-unit properties
\beq
m\cdot (\e\otimes \I)F=I=m\cdot(\I\otimes \e) F.\label{e(f)=1}
\eeq
We set
\bea
&&\D_F(a)=F\D(a)F^{-1},~~~\forall a\in A,\label{twisted-d}\\
&&\Phi_F=F_{12}(\D\otimes
    \I)F\cdot\Phi\cdot(\I\otimes\D)F^{-1}F_{23}^{-1}.\label{twisted-phi}
\eea 
The element $F$ is referred to as
a  twistor.
\begin{Theorem}\label{t-quasi-hopf}:
If $(A,\D,\e,\Phi)$ is a quasi-bialgebra then $(A,\D_F,\e,\Phi_F)$ 
is also a quasi-bialgebra.  
If $(A,\D,\e,\Phi,S,\alpha,\beta)$ is a quasi-Hopf algebra then 
$(A,\D_F,\e,\Phi_F,S_F,\a_F,\b_F)$ is also a quasi-Hopf algebra where  
$\a_F,\b_F, S_F$ are given by
\beq
S_F=S,~~~\a_F=m\cdot(I\otimes\a)(S\otimes \I)F^{-1},~~~
       \b_F=m\cdot(I\otimes\b)(\I\otimes S)F. \label{twisted-s-ab}
\eeq
If $(A,\D,\e,\Phi,\R)$ is a  quasi-triangular quasi-bialgebra 
then $(A, \D_F, \e, \Phi_F, \R_F)$
is a quasi-triangular quasi-bialgebra with 
$R_F$ given by
\beq
\R_F=F^T \R F^{-1}.\label{twisted-R}
\eeq
\end{Theorem}
The above shows that the categories of quasi-bialgebras, 
quasi-Hopf algebras and
quasi-triangular quasi-bialgebras are all invariant under twisting.
This leads to the following result \cite{cp} 

\begin{Theorem}\label{result1} 
If $A$ is a quasi-bialgebra then the category $\mathbf{mod}(A)$ of
finite-dimensional $A$-modules is a monoidal category. If $A$ is a
quasi-Hopf algebra then $\mathbf{mod}(A)$ is a rigid, monoidal category.
If $A$ is a quasi-triangular quasi-bialgebra then $\mathbf{mod}(A)$
is a braided, monoidal category. 
Suppose that two quasi-bialgebras $A$ and $B$ are twist equivalent. Then
$\mathbf{mod}(A)$ and $\mathbf{mod}(B)$ are equivalent as monoidal
categories. If $A$ and $B$ are are equivalent as quasi-Hopf algebras, then
$\mathbf{mod}(A)$ and $\mathbf{mod}(B)$ are equivalent as rigid, monoidal
categories. If $A$ and $B$ are equivalent as quasi-triangular
quasi-bialgebras, then $\mathbf{mod}(A)$
and $\mathbf{mod}(B)$ are equivalent as braided, monoidal categories. 
\end {Theorem} 

It is useful to give an outline of the above result. For every
$A$-module $U$ let $\pi_U: A \rightarrow {\rm End}\,U$ be the
representation afforded by $U$. By defining  
\beq a_{U,V,W}=(\pi_U\otimes \pi_V\otimes\pi_W)\Phi \label{iso} \eeq  
(\ref{quasi-bi1}) ensures each of the $a_{U,V,W}$ are $A$-module
isomorphisms, for all $A$-modules $U,V$, and $W$. The axiom
(\ref{pent}) ensures commutativity of the
pentagon diagram is satisfied. Since 
the trivial one-dimensional module is defined by
the action of the co-unit $\e$, and (\ref{quasi-bi3}) is
satisfied, the natural isomorphisms $\rho_U$ and $\lambda_U$ are 
$A$-module isomorphisms for any $A$-module $U$. Finally, the axiom
(\ref{quasi-bi4}) ensures that the diagram of Definition \ref{upmc}
commutes.

The existence of the antipode for quasi-Hopf algebras allows for 
the definition of dual modules. This guarantees that the category of
finite-dimensional quasi-Hopf algebra modules has a rigid monoidal
structure. To make this explicit, 
for a basis $\{ u_i \}$ of $U$ 
we define the dual basis $\{ u^i \}$ of
$U^*$ such that for all $u\in U$, $u=\sum_i u^i(u) u_i.$ The
natural pairing $<u^i,u_j>=u^i(u_j) = \delta^i_j$, along with the
antipode, gives the action of the elements of $A$ on $U^*$ by
$$
<a.u^*, u> = <u^*, S(a).u>, ~~~\forall a\in A, u^*\in U^*, u\in U.
$$
We define $\ev_U$ and $\coev_U$ as follows: 
$$
\ev_U(u^*\tp u) = <u^*,\a.u>, \ ~~~\coev_U(1) = \sum_i \b.u_i\tp u^i.
$$
Axioms (\ref{quasi-hopf1}) and (\ref{quasi-hopf2}) ensure $\ev_U$
and $\coev_U$ are $A$-module homomorphisms for any 
finite-dimensional $A$-module $U$. Commutativity of the two diagrams in
our discussion on dual modules and rigidity in 
Section \ref{cats} is guaranteed by (\ref{quasi-hopf3}) and
(\ref{quasi-hopf4}).

To see that the category of finite-dimensional representations of a
quasi-triangular quasi-bialgebra is braided, we first introduce the
permutation operator $P:U\tp V\rar V\tp U$ for any $A$-modules $U$ and $V$
via $P(u\tp v)=v\tp u$ for all $u\in U, v\in V$. We then define the braiding
morphism  
$\sigma_{U,V}:U\tp V\rar V\tp U$ as 
$$ \s_{U,V} = P\cdot (\pi_U\tp \pi_V) \R. $$
It is convenient to express  
the action of $\Phi_{132}$ etc. on the tensor product
of modules in terms of the action of $\Phi$ and $P$. 
For example, (we drop the symbols $\pi_U$ denoting the representations for 
notational ease) 
\bea
\Phi_{132} & = & (\I\tp P)\cdot\Phi\cdot(\I\tp P) \no\\ 
\Rightarrow \Phi & = & (\I\tp P)\cdot\Phi_{132}\cdot(\I\tp P), \label{eg1} \\ 
\Phi_{231} & = &
(\I\tp P)\cdot(P\tp \I)\cdot\Phi\cdot(P\tp \I)\cdot(\I\tp P) \no \\
\Rightarrow \Phi & = & (P\tp \I)\cdot(\I\tp P)\cdot\Phi_{231}\cdot(\I\tp
P)\cdot (P\tp \I), \label{eg2} \eea
and so on. Also note that identities between the actions of $\R$
and $P$ exist such as 
\beq
(\I\tp P)\cdot\R_{13} = (R\tp \I)\cdot(\I\tp P).  \label{eg3} \eeq
All of these identities can be easily established by direct calculation. 
Keeping in mind the fact 
that $P^2=\I\tp \I$, by considering (\ref{d1r}) and (\ref{1dr}) as acting on 
the three-fold tensor product space of modules, we can express both
of these equations purely in terms of $\Phi$ and $\s$.  To
demonstrate, we look in detail at (\ref{d1r}).  First express this equation as
$$ \Phi_{231}\cdot(\D\tp\I)\R\cdot \Phi_{123} =
\R_{13}\cdot\Phi_{132}\cdot\R_{23}.  $$
Multiplying both sides on the left by $(P\tp \I)(\I\tp P)$ 
leads to the equality 
\bea && (P\tp \I)\cdot(\I\tp P)\cdot\Phi_{231}\cdot(\I\tp P)\cdot(P\tp \I)
\cdot(P\tp \I)\cdot(\I\tp P)\cdot(\D\tp\I)\R\cdot\Phi_{123} \qquad \no\\
&& \qquad = \ (P\tp \I)\cdot(\I\tp P)\cdot \R_{13}\cdot\Phi_{132}\cdot(\I\tp P)
\cdot(\I\tp P)\cdot \R_{23}.  \no \eea 
Applying (\ref{eg1}), (\ref{eg2}) and (\ref{eg3}) and simplifying
the expressions leads to $$ \Phi\cdot(P\tp \I)\cdot(\I\tp P)\cdot(\D\tp\I)\R
\cdot\Phi =(\s\tp \I)\cdot\Phi\cdot(\I\tp \s).  $$ 
We consider the composite $(P\tp \I)\cdot(\I\tp P)\cdot(\D\tp\I)\R$
to be the braiding isomorphism $\s_{(U\tp V),W}$.  Since we are defining 
the associativity isomorphisms $a_{U,V,W}$ as the image of $\Phi$
through (\ref{iso}), this last
equality exactly describes the diagram (ii) in Definition \ref{bmc}.
The same calculation can be applied to (\ref{1dr}) to obtain commutativity
of the diagram (i) of Definition \ref{bmc}.
It should be noted that the axiom (\ref{dr=rd}) ensures
that $\s$ is an $A$-module homomorphism, and diagram (iii) of Definition
\ref{bmc} is trivially satisfied in the monoidal case. 
Finally, in cases where $A$ and $B$ are twist equivalent the relations 
(\ref{twisted-d},\ref{twisted-phi},\ref{twisted-s-ab},\ref{twisted-R}) 
provide isomorphisms between $\mathbf{mod}(A)$ and
$\mathbf{mod}(B)$. 

\section{Twining and universal enveloping algebras of Lie algebras}

In order to produce non-trivial braided pre-monoidal categories  
we introduce a deformation, referred to as twining, 
through the use of the Casimir invariants of
the algebra. Given a quasi-triangular quasi-bialgebra which possesses
a Casimir invariant $K$,  let us define for a fixed but arbitrary
$\gamma\in{\mathbb C}$ 
\bea   
\tR&=&\gamma^{K\otimes K}\cdot\R=\R\cdot\gamma^{K\otimes K} \no \\
\tP&=&\Phi\cdot\gamma^\kappa \label{twining} \eea 
where
\beq \kappa=
{K\otimes(I\otimes K+K \otimes I-\D(K))}. \label{kappa} \eeq        
The following relations hold 
\bea
&&(\I\otimes \D)\D(a)=\tP^{-1}(\D\otimes \I)\D(a)\tP~~~~\forall a\in
A,\no\\
&&\tR\D(a)=\D^T(a)\tR,~~~~\forall a\in A,\no\\
&&(\D\otimes
\I)\tR= \tP^{-1}_{231}\tR_{13}\tP_{132}\tR_{23}\tP^{-1}_{123},
   \no \\
   &&(\I\otimes \D)\tR=\tP_{312}\tR_{13}\tP^{-1}_{213}\tR_{12}\tP_{123}
   (\gamma^{2\kappa})^{-1}_{123} \label{coeff} \eea  \\
and we define 
\beq \xi=(\D\otimes \I\otimes \I)\tP^{-1}\cdot(\tP\otimes I)\cdot(\I\otimes
\D\otimes \I)\tP\cdot(I\otimes\tP)\cdot(\I\otimes \I\otimes \D)\tP^{-1}.
\label{xi} \eeq 
Note that the quasi-Yang-Baxter equation (\ref{qybe}) is still satisfied. 

{}From the above relations we immediately see that the category 
of $A$-modules with 
$$a_{U,V,W}=\l(\pi_U\tp\pi_V\tp \pi_W\r)\tP, $$
denoted
${\mathbf{mod}}_K(A)$, is a 
pre-monoidal category since in this case $\tP$  
generally fails to satisfy the pentagon relation (\ref{pent}); i.e. the
representation 
$$q_{U,V,W,Z}=(\pi_U\tp\pi_V\tp\pi_W\tp\pi_Z)\xi$$ 
is not the identity. 
In the language of \cite{majid} we can identify (\ref{twining}) as a 3-co-chain
with non-trivial co-boundary, so $\tP$ is not a 3-co-cycle. 

However, $\tR$ 
cannot be used to construct a braided, pre-monoidal category of
$A$-modules, 
due to the term $\gamma^{2\kappa}$ in (\ref{coeff}) which violates the hexagon
condition (i) of Definition \ref{bmc}. On the
other hand, it may well be possible that for suitably chosen $\gamma$
there is a subcategory of  
${\mathbf{mod}}_K(A)$ for which $\gamma^{2\kappa}=1$ when restricted to this
subcategory. In this case, such a subcategory may acquire the
structure of a rigid, braided, pre-monoidal category. Below we
demonstrate that this is indeed possible when $A$ is the universal
enveloping algebra $U(g)$ of a simple Lie algebra $g$. 
Moreover, we will see that
the subcategory is the full category 
${\mathbf{mod}}_K(A)$ containing all the finite-dimensional irreducible 
$U(g)$-modules. Finally, we make mention of the fact that twisting and twining
are not commutative unless 
$$[\D(K),\,F]=0, $$      
as can be seen from (\ref{twisted-phi}). 
Thus combinations of twisting and twining can be 
performed to produce many examples of pre-monoidal categories. It can be 
verified that under twisting $\xi$ transforms as 
$$\xi_F=(F\tp F)\cdot(\D\tp \D)F\cdot\xi\cdot(\D\tp \D)F^{-1}\cdot(F^{-1}
\tp F^{-1}).$$

Recall that, for given Cartan matrix $A=(a_{ij})$ of rank $n$, 
the simple Lie algebras can be defined in terms 
of simple generators $h_i,\,e_i,\,f_i,\,i=1,...,n$ 
subject to the relations 
\bea &&[h_i,\,h_j]=0, \no \\
&&[h_i,\,e_j]=a_{ij}e_j \no \\
&&[h_i,\,f_j]=-a_{ij}f_j \no \\
&&[e_i,\,f_j]=\delta_{ij}h_i \no \\
&&({\rm ad}\,e_i)^{1-a_{ij}}e_j=({\rm ad}\,f_i)^{1-a_{ij}}f_j=0,~~~~~
{\rm for}~i\neq j\no \eea 
where ad refers to the adjoint action. The universal enveloping algebra
$U(g)$ of a Lie algebra $g$ acquires the structure of a quasi-bialgebra with
the mappings 
\bea 
&\e(I)=1,~~~~~~~~~~~~\e(x)=0,\,~~~~~~~~&\forall x\in g \no \\
&S(I)=I,~~~~~~~~~~~S(x)=-x,\,~~~~~~~~&\forall x\in g \no \\
&\D(I)=I\otimes I,~~~\D(x)=I\otimes x+x\otimes I,\,~~&\forall x\in g  
\no \eea 
which are extended to all of $U(g)$ such that $\e$ and $\D$ are algebra
homomorphisms and $S$ is an anti-automorphism. 
It is easily checked that $\D$ is co-associative; i.e. 
$$(\I\otimes\D)\D(x)=(\D\otimes \I)\D(x)~~~~~~\forall x\in U(g). $$ 
This means that we can
take $\Phi=I\otimes I\otimes I$ for the co-associator of $U(g)$ and 
$\alpha=\beta=I$.

The following is a well known theorem from the representation theory of
Lie algebras. 
\begin{Theorem} \label{hw} 
All the finite-dimensional modules of a simple Lie algebra 
$g$ are completely reducible.  Every finite-dimensional 
irreducible $g$-module $V(\Lambda)$ is uniquely characterised by the 
highest weight $\Lambda=(\Lambda_1,...,\Lambda_n)$ associated with the 
unique highest weight vector $v^+\in\,V$ satisfying
$$e_iv^+=0,~~~h_iv^+=\Lambda_iv^+,~~~\,\Lambda_i\in{\mathbb
C},~~~~i=1,...,n.  $$ 
A necessary and sufficient condition for $V(\Lambda)$ to be
finite-dimensional is that each $\Lambda_i$ is a non-negative integer. 
\end{Theorem} 

There are several works devoted to the construction of Casimir
invariants for Lie algebras. Generally, such invariants are polynomial
functions of the Lie algebra generators, and consequently the
eigenvalues of a Casimir invariant acting on an irreducible
finite-dimensional module $V(\Lambda)$ are polynomial functions of the
$\Lambda_i$ (e.g. see \cite{mark}). 
Thus, Theorem \ref{hw} assures that we may find a 
Casimir invariant $K$ which, with appropriate normalisation, 
takes integer values in all irreducible
finite-dimensional representations. We also impose that  $\e(K)=0$, so
that (\ref{quasi-bi4}) remains true (note this rules out the
possibility of choosing $K=I$).
We call such Casimir invariants {\em
admissible}, the set of which forms a ring. 
We use the notation $\chi_{\Lambda}(K)$ to denote the eigenvalue of $K$
on $V(\Lambda)$. Now choosing $\gamma=-1$ in (\ref{twining}) 
permits us to apply twining to construct a  
symmetric category of $U(g)$-modules which is generically pre-monoidal.
In particular, for $\R=I\otimes I$ and 
$\Phi=I\otimes I \otimes I$ we find that 
\bea 
\tR&=&(-1)^{K\otimes K} \label{choice1} \\
\tP&=&(-1)^{K\tp (I\tp K +K\tp I-\D(K))} \label{choice2} \\
\xi&=&(-1)^{(I\tp K +K\tp I-\D(K))\tp (I\tp K +K\tp I-\D(K))} 
\label{choice3} \eea 

Our final requirement is to extend the above constructions to the case
of finite-dimensional semi-simple Lie algebras. 
A straightforward way to do this is 
the following inductive method, which will be employed in an example
below. Supposing $g=h\oplus k$ where $h$ is simple  
we impose that each state in a $U(h)$-module is a
singlet state with respect to $U(k)$, so the action of $U(k)$ on a
$U(h)$-module is given by the co-unit of $U(k)$, and vice versa.  
This definition of the action ensures that if $K^h$ and $K^k$ are
admissible Casimir invariants for $h$ and $k$ respectively then 
$$K^{h\oplus k}=K^h+K^k$$
is admissible for $h\oplus k$, with eigenvalue on the $(h\oplus k)$-module  
\beq V(\Lambda|\Lambda')=V^h(\Lambda)\otimes V^k(\Lambda') \label{object} \eeq 
given by 
\beq \chi_{K^{h\oplus k}}(\Lambda|\Lambda')=\chi_{K^h}(\Lambda)
+\chi_{K^k}(\Lambda'). \label{eig} \eeq  
In this way a twining defined on $k$ can be extended to a twining
on $g$.      

\begin{Proposition} 
For any finite-dimensional 
semi-simple Lie algebra $g$ with Casimir invariant $K$ taking integer
values in all irreducible finite-dimensional representations, the
category ${\mathbf{mod}}_K(U(g))$ of $U(g)$-modules is a braided,  
pre-monoidal category with 
\bea a_{U,V,W}&=&(\pi_U\otimes\pi_V\otimes\pi_W)\tP \no \\    
q_{U,V,W,Z}&=&(\pi_U\otimes\pi_V\otimes\pi_W\otimes\pi_Z)\xi \no \\
\sigma_{U,V}&=&P(\pi_U\otimes\pi_V)\tR. \no \eea    
For the particular choices (\ref{choice1},\ref{choice2},\ref{choice3})
the category is symmetric. If moreover $S(K)=\pm K$, the category is
symmetric and rigid. 
\end{Proposition}
The 
proof is essentially the same as that used to prove Theorem \ref{result1}.
By construction, the isomorphisms 
$q_{U,V,W,Z}$ 
are not generically the identity isomorphism, and so the category is not
monoidal in general. 
Symmetry in the case of (\ref{choice1},\ref{choice2},\ref{choice3}) 
follows from the definition of $\sigma_{U,V}$ in terms of $\tR$.  
From (\ref{choice2}) we find   
\bea 
&& m\cdot (m\otimes \I)\cdot (\I\otimes S\otimes
     \I)\tP^{-1}=(-1)^{-K(K+S(K))}  ,\no \\
     && m\cdot(m\otimes \I)\cdot(S \otimes \I\otimes S)\tP
     =(-1)^{S(K)(K+S(K))} \no 
\eea 
so $S(K)=\pm K$ then ensures rigidity. Note that this condition is
sufficient, but by no means necessary. 
Diagram (iii) of Definition \ref{bmc} holds due to the commutativity of
$K$ with all elements of $U(g)$. 
It is important to make it clear that in the case of a semi-simple Lie
algebra the objects of ${\mathbf{mod}}_K(U(g))$ are the modules of the
form (\ref{object}), not the individual tensor components $V^h(\Lambda)$ and
$V^k(\Lambda')$. 

The above constructions allow a principle of confinement and exclusion to
be formulated for indistinguishable particles. In the following we
restrict our focus to the cases governed by
(\ref{choice1},\ref{choice2},\ref{choice3}). 
Given an irreducible
finite-dimensional $U(g)$-module 
$V\equiv V(\Lambda)\in\,{\mathbf{mod}}_K(U(g))$ 
we can assign $V$ to be the space of
states for a {\em fundamental particle} with parity
\beq p(\Lambda)=\chi_K(\Lambda)({\rm mod}\,2). \label{bf} \eeq  
This defines a statistic for the fundamental particle. 
Particles with parity zero are said to be bosons while those with parity
one are fermions. It is important to stress that hereafter the terms
boson and fermion are in reference to (\ref{bf}) which is more general
than  the usual understanding of these words. The significance of this 
difference will be
made apparent in later examples. 
Let ${\mathbf{mod}}^V_K(U(g))\subset{\mathbf{mod}}_K(U(g))$ 
be the subcategory generated by $V$ and the identity module. 
A state $\l|\Psi\r>$ 
of $N$ fundamental 
particles is a vector contained in an $N$-fold tensor
product space of $V$, called an {\em ensemble}. Each ensemble  
is an object of ${\mathbf{mod}}^V_K(U(g))$. 
For all $U\in {\mathbf{mod}}^V_K(U(g))$ the isomorphisms $a_{V,V,V}$
allow us to construct a recoupling morphism 
${\mathbf{rec}}^U: {\mathbf{mod}}^V_K(U(g))
\rightarrow {\mathbf{mod}}^V_K(U(g))$ such that $U$ is 
mapped to the {\em standard presentation}
$${\mathbf{rec}}^U(U)=((...(((V\tp V)\tp V)\tp V)...)\tp V). $$
We can extend ${\mathbf{rec}}^U$ to a morphism on states $\l|\Psi\r>\in
U$ in an obvious way. Note that ${\mathbf{rec}}^U$ is not unique. For
example, in the case $U=(V\tp V)\tp (V\tp V)$ the following 
\bea 
{\mathbf{rec}}_1^U&=&a^{}_{(V\tp V),V,V} \label{rec1} \\
{\mathbf{rec}}^U_2&=&(a^{}_{V,V,V}\tp \I)a^{}_{V,(V\tp V),V}(\I\tp
a^{}_{V,V,V})a^{-1}_{V,V,(V\tp V)} \label{rec2} \eea 
are both examples of recoupling morphisms, which are not necessarily
equal since the pentagon axiom does not hold in general. 

Following \cite{ac,lgz}, 
define $\D^{(n)}: U(g) \rightarrow U(g)^{\tp (n+1)}$ through
$$\D^{(n)}=\l(\D\tp \I^{\tp(n-1)}\r)\cdot \D^{(n-1)}$$
with $\D^{(0)}=\I$, set 
$$\tP_{(1)}=\I^{\tp N},~~~\tP_{(i)}=(\D^{(i-2)}\tp \I
\tp \I)\tP\tp \I^{(N-i-1)},\ i=2,\ldots, N-1,$$ 
and let
\bea 
\sigma_{(i)}&=&\pi_V^{\tp N}(\tP_{(i)})\cdot \l(\I^{\tp (i-1)}\tp
\sigma_{V,V}\tp \I^{\tp (N-i-1)}\r)\cdot \pi_V^{\tp N}(\tP_{(i)}^{-1})
\no \\
&=&\I^{\tp (i-1)}\tp
\sigma_{V,V}\tp \I^{\tp (N-i-1)}. \no \eea 
The operators $\sigma_{(i)}$ provide a representation of the symmetric group
$S_N$ with the relations 
\bea 
&&\sigma_{(i)}\sigma_{(i+1)}\sigma_{(i)}=\sigma_{(i+1)}\sigma_{(i)}
\sigma_{(i+1)}, \no \\
&&\sigma_{(i)}\sigma_{(j)}=\sigma_{(j)}\sigma_{(i)}
~~~~~~~~~~~~~~~~~~~~~~j\neq i\pm 1, \no
\\
&&\sigma_{(i)}^2=\I^{\tp N} \no \eea 
on any 
${\mathbf{rec}}^U(U)\in{{\mathbf{mod}}}^V_K(U(g))$. 

Since the particles are indistinguishable, we require that observable 
states are invariant with
respect to the action of the symmetric group. This provides the
underlying principle for confinement and exclusion.
Specifically, given a state $\l|\Psi\r>\in U
\tp U$ this means 
$$\sigma_{U,U}\l|\Psi\r>=\l|\Psi\r>. $$ 
We also require that the extension to
coupled states is consistent. In general, we have from Definition
\ref{bmc}
\bea 
\sigma_{V,(W\tp Z)}=a_{W,Z,V}(\I\tp \sigma_{V,Z})a^{-1}_{W,V,Z}
(\sigma_{V,W}\tp \I)a_{V,W,Z} \no \\
\sigma_{(V\tp W),Z}=a^{-1}_{Z,V,W}(\sigma_{V,Z}\tp \I)a_{V,Z,W}
(\I\tp \sigma_{W,Z})a^{-1}_{V,W,Z}. \no \eea   
Combining these expressions and  specialising to the case $U=V=W=Z$ 
yields
\bea \sigma^{}_{(U\tp U),(U\tp U)}&=&a^{-1}_{(U \tp U),U,U}
(a^{}_{U,U,U}\tp \I)(\I\tp \sigma^{}_{U,U} \tp \I)(a^{-1}_{U,U,U}\tp \I) 
 \no \\
 &&~~~\times (\sigma^{}_{U,U}\tp \I\tp \I)(a^{}_{U,U,U}\tp \I)
 a^{}_{U,(U\tp U),U} \no \\
&&~~~\times 
(\I\tp a^{}_{U,U,U})(\I\tp \I\tp\sigma^{}_{U,U})(\I\tp a^{-1}_{U,U,U}) \no \\
&&~~~\times (\I\tp \sigma^{}_{U,U}\tp \I)(\I\tp a^{}_{U,U,U})
a^{-1}_{U,U,(U\tp U)}.  \no \eea  
Applying this operator to a state $\l|\Psi\r>\in (U\tp U)\tp (U\tp U)=Y$ 
and assuming invariance with respect to the symmetric group acting on
$Y$ we find 
\bea 
&&\sigma^{}_{(U\tp U),(U\tp U)}\l|\Psi\r>   \no \\ 
&&~~~~=\l[{\mathbf{rec}}^Y_1\r]^{-1}\circ \l[\sigma_{(1)}\sigma_{(2)} 
{\mathbf{rec}}^Y_2\r]\circ 
\l[{\mathbf{rec}}^Y_1\r]^{-1}\no \\
&&~~~~~~~~~~~~~\circ\l[\sigma_{(3)}{\mathbf{rec}}^Y_1\r] 
\circ \l[{\mathbf{rec}}^Y_2\r]^{-1}\circ\sigma_{(2)}
{\mathbf{rec}}^Y_2
\l(\l|\Psi\r>\r) \no \\
&&~~~~=\l[{\mathbf{rec}}^Y_1\r]^{-1}\circ{\mathbf{rec}}^Y_2
\l(\l|\Psi\r>\r) \no \\
&&~~~~=a^{-1}_{(U \tp U),U,U}(a^{}_{U,U,U}\tp \I)a^{}_{U,(U\tp U),U}(\I \tp
a^{}_{U,U,U}) a^{-1}_{U,U,(U\tp U)}\l|\Psi\r> \no \\
&&~~~~=q^{}_{U,U,U,U}\l|\Psi\r> \no  
\eea 
with ${\mathbf{rec}}^W_1$ and ${\mathbf{rec}}^W_2$ given by
(\ref{rec1},\ref{rec2}). 
Invariance with respect to the symmetric group for 
coupled states imposes that observable states 
are restricted to a
monoidal subcategory where $q^{}_{U,U,U,U}=\I\tp \I\tp\I\tp \I$. Therefore
let 
${\overline{\mathbf{mod}}}^V_K(U(g))$ denote the
{\em maximal} monoidal subcategory of ${\mathbf{mod}}^V_K(U(g))$. 

\begin{Proposition} \label{observables} 
{\em Principle of observable states for many indistinguishable
fundamental particles}: Given a $U(g)$-module $V$ which characterises the
state space of a fundamental particle,  an $N$-particle
state  $\l|\Psi\r>\in U$, where
$U\in {\mathbf{mod}}^V_K(U(g))$, is
said to be {\em observable} if $U\in {\overline{\mathbf{mod}}}^V_K(U(g))$ and 
$$\sigma_{(i)}\cdot{\mathbf{rec}}^U\l(\l|\Psi\r>\r)=
{\mathbf{rec}}^U\l(\l|\Psi\r>\r),~~~ \forall \,i=1,...,N-1. $$
An observable state is said to be {\em elementary} if it cannot be
expressed as a (non-trivial) 
linear combination of tensor products of observable states. 
A subspace of
elementary states which is invariant and irreducible under $U(g)$ is
said to be the state space for an elementary particle. 
\end{Proposition} 

The above proposition combines the notions of exclusion and confinement
of states as a consequence of symmetry rather than dynamics. 
First note that ${\overline{\mathbf{mod}}}^V_K(U(g))$ is always 
non-empty, since it at least contains the identity module. The fact that
$\e(K)=0$ shows that the identity module is always bosonic, and can be
identified as the vacuum. 
Any state $\l|\Psi\r>\in {\overline{\mathbf{mod}}}^V_K(U(g))$ 
which is not invariant
under the action of $\sigma_{(i)},\,\forall\,i=1,...,N-1$ is excluded. 
A consequence of this definition is that an observable state is either
completely symmetric (when the statistic is bosonic) 
or completely antisymmetric (when the statistic is fermionic) 
with respect to the interchange of {\em
any} two fundamental particles, since any such interchange can be 
expressed as a product of $\sigma_{(i)}$ with an odd number of terms. 
A set of states $\{\l|\Psi_j\r>\}\notin
{\overline{\mathbf{mod}}}^V_K(U(g))$ are not observable but there may
exist a set $\{\l|\theta_k\r>\}$ such that 
$\sum_{j,k}\alpha_{jk}\l|\Psi_j\r>\otimes \l|\theta_k\r> \in
{\overline{\mathbf{mod}}}^V_K(U(g))$ is observable for appropriate 
$\alpha_{jk}\in{\mathbb{C}}$. In this sense
the states $\{\l|\Psi_j\r>\}$ are confined. 
Within this context confinement does not depend on the existence of
an attractive confining force, in the same way that the familiar concept of
exclusion of fermions is not dependent on the existence of a repulsive
force that prohibits fermions from occupying the same state. 
Both are a consequence of a prescribed statistic for a 
given symmetry. Note finally that our definition for an elementary  
particle is in no way exhaustive, because there is no reference to the
role played by antiparticles. The fact that 
${{\mathbf{mod}}}_K(U(g))$
is rigid (for suitable $K$) 
is significant in that for every module denoting a fundamental  
particle there exists a dual module for the antiparticle, and similarly
for elementary particles. In such a case one may include an additional
$u(1)$ symmetry which distinguishes particles and antiparticles. 
This will not be considered here, but can be achieved following the
procedure given in \cite{ijl}. 

Proposition \ref{observables}, while in the same spirit as the
confinement and exclusion principles given in 
\cite{joyce-s}, is formulated in a different manner. In
\cite{joyce-s} the criteria for exclusion and confinement are given with
respect to a specific statistic defined for any simple Lie algebra $g$.
The formulation here allows  a range of different statistics to be chosen for
$g$ through
different choices of the Casimir invariant. 
Through this approach we will recover  
the results given in \cite{joyce-s} for the cases of $su(2)$ and $su(3)$. 

Hereafter we will only concern ourselves with the $su(n)$ algebras. 
For these algebras the defining representation $\pi:su(n)\rightarrow
\,{\rm End}\,{\mathbb C}^n$ is given by 
$$\pi(e_i)=E^i_{i+1},~~~~\pi(f_i)=E_i^{i+1},~~~~\pi(h_i)=E^i_i-E^{i+1}_{i+1}
 $$ 
where $E^i_j$ denotes the matrix with 1 in the $(i,j)$ entry and zeros
elsewhere. 
For this class of algebras it is useful to introduce the {\it arity} of
a finite-dimensional irreducible module 
\cite{joyce-v}. Let the defining module $V(1,0,...,0)$ have arity one. 
If 
$$V(\Lambda)\subset V(1,0,...,0)^{\otimes m}$$   
we say that $\Lambda$ has arity $ar(\Lambda)=m\,({\rm mod}\, n)$.  
It follows that for 
$$V(\Lambda)\subset V(\Lambda')\otimes V(\Lambda'')$$ 
then 
\beq ar(\Lambda)=\left(ar(\Lambda')+ar(\Lambda'')\right)({\rm mod}\, n). 
\label{aritysum} \eeq  
The notion of arity permits us to define equivalence classes on the
modules. Specifically, 
$$\Lambda\sim\Lambda' ~~~~{\rm if~ and~ only~ if}
~~~~ ar(\Lambda)=ar(\Lambda').$$ 
Hereafter all subscripts on component isomorphisms refer to the arity
associated with each equivalence class. In the case of $su(2)$, arity
corresponds to the usual notion of parity. For $su(3)$, it is often
referred to as triality \cite{gn,hn,pol}.

Finally, suppose there exists a Casimir invariant $C\in U(su(n))$ such that 
$$ar(\Lambda)=\chi_C(\Lambda)({\rm mod}\,n). $$   
Setting $\mu=\exp(2\pi i/n)$ it follows from (\ref{aritysum}) that 
\bea 
\mu^{\D(C)-I\tp C-C\tp I}\l(V(\Lambda)\otimes V(\Lambda')\r)&=& 
\mu^{ar(\Lambda\otimes \Lambda')-ar(\Lambda)-ar(\Lambda')}
\l(V(\Lambda)\otimes V(\Lambda')\r)\no \\
&=&
V(\Lambda)\otimes V(\Lambda') \label{identity}. \eea  

\section{The $su(2)$ case} 

As the first illustration of how the theory applies we now look at the
$su(2)$ case which reproduces 
the spin statistics theorem. 
It is well known that the finite-dimensional $su(2)$ modules, labelled
by a non-negative integer $\Lambda$, have dimension $\Lambda+1$. The
highest weight label $\Lambda$ is related to the more familiar spin 
label $s$ through 
$$s=\frac{\Lambda}{2}.$$ 
The spin statistics theorem asserts that the even-dimensional  modules describe
fermionic states while the odd-dimensional modules  are bosonic. 

The algebra $U(su(2))$ has one independent Casimir invariant, viz. 
the second order invariant 
$$C_2=2ef+2fe+h^2.$$ 
We begin by noting that 
the second order invariant takes the eigenvalue 
$$\chi_{\Lambda}(C_2)=\Lambda(\Lambda+2) $$ 
when acting on the finite-dimensional irreducible module $V(\Lambda)$. 
It is apparent that if $\Lambda$ is even (resp. odd), then 
$\chi_{\Lambda}(C_2)$ is even (resp. odd). 
In this case the category ${\mathbf {mod}}_{C_2}(U(su(2)))$ degenerates to a
monoidal category. Because of (\ref{identity}) we conclude that 
\bea &&(\pi_{U}\otimes \pi_{V}\tp \pi_W)(-1)^{C_2\tp(\D(C_2)-I\otimes C_2 
-C_2\otimes I)} \no \\
&&~~~~=(\pi_V\tp \pi_W)(-1)^{\chi_U(C_2)(\D(C_2)-I\tp C_2-C_2\tp I)}\no \\
&&~~~~=\I\otimes \I\tp \I, ~~~~\forall\,U,\,V,\,W \in\,
{\mathbf {mod}}_{C_2}(U(su(2)))  \no \eea  
where the last line follows because $\chi_U(C_2)$, the eigenvalue of 
$C_2$ on $U$, is an integer. 
Consequently the pentagon axiom is
satisfied and there is no confinement of states. 

The  braiding morphism defined by
$$\sigma_{U, V}=P (\pi_{U}\otimes
\pi_{V})(-1)^{C_2\otimes C_2}$$
yields a symmetric, monoidal category of finite-dimensional
modules for $U(su(2))$. The component isomorphisms are found to be
$$\tR_{0,0}=\tR_{0,1}=\tR_{1,0}=-\tR_{1,1}=I\tp I. $$
This exactly reproduces the spin statistics theorem.
Observable states of coupled indistinguishable fundamental particles are
completely symmetric if the particles are bosons and completely
antisymmetric if they are fermions. 
For pedagogical purposes, let us elaborate on the case of a 
fundamental spin-1/2
particle with no other degrees of freedom. 
Since ${\mathbf {mod}}_{C_2}(U(su(2)))$ is monoidal,    
the spin-1/2 particle, in either spin state, is observable, and thus 
a spin-1/2 particle is elementary. Two spin-1/2 particles 
can be coupled provided the state is antisymmetric, which gives the
pair singlet. This particle is not elementary by our
definition. Exclusion
prohibits any higher order states being observable. Thus there are four
independent observable states; 
the vacuum, the spin up state, the spin down state
and the spin singlet. For a many-body system with additional quantum
state labels, which afford many degrees of freedom, 
these observable states may provide a local space of states.
Examples in the area of condensed matter physics 
are the Hubbard model, where there is a quantum
state label for the sites of the lattice, and the BCS model, where
there is an additional label for the single particle energy levels.

\section{The $su(3)$ case}\label{colour} 

Here we extend the above analysis to the case of $su(3)$ and will show
that in this instance the boson-fermion statistic 
leads to a symmetric  pre-monoidal category of
finite-dimensional modules. Applying the $su(3)$ algebra as the symmetry
algebra for colour charges which mediate the strong interactions in the
Standard Model, we
will see that this construction leads to colour confinement. 
The $U(su(3))$ algebra has two Casimir invariants, explicitly given by 
\cite{rp} 
\bea 
C_2&=&(h_1^2+h_2^2+h_1h_2) +\frac{3}{2} (e_1f_1+f_1e_1+e_2f_2+f_2e_2 
-e_{12}f_{12}-f_{12}e_{12}) \no \\
C_3&=&(2h_2^3-2h_1^3+3h_1h_2^2-3h_1^2h_2) \no \\ 
&&~~-\frac{3}{2}(s(h_1+2h_2,e_1,f_1)
-s(2h_1+h_2,e_2,f_2)-s(h_1-h_2,e_{12},f_{12})
)\no \\
&&~~+\frac{9}{2}(s(e_1,e_2,f_{12})-s(f_1,f_2,e_{12})) \no \eea   
where 
\bea 
&&e_{12}=e_1e_2-e_2e_1, \no \\
&&f_{12}=f_1f_2-f_2f_1 \no \eea 
and 
$$s(a,b,c)=abc+bca+cab+acb+cba+bac. $$ 
In terms of the highest weight labels $\Lambda=(\Lambda_1,\,\Lambda_2)$
the eigenvalues of the Casimir invariants on the irreducible
finite-dimensional module $V(\Lambda)$ read \cite{rp} 
\bea 
\chi_{\Lambda}(C_2)&=&(\Lambda_1^2+\Lambda_2^2+\Lambda_1\Lambda_2)+
3(\Lambda_1+\Lambda_2) \no \\
\chi_{\Lambda}(C_3)&=&(\Lambda_2-\Lambda_1)(2\Lambda_1+\Lambda_2+3)
(\Lambda_1+2\Lambda_2+3).  \eea      
It is worth noting that the dimensionality of $V(\Lambda)$ is 
$${\rm dim}(V(\Lambda))=\frac{1}{2}(\Lambda_1+1)(\Lambda_2+1)
(\Lambda_1+\Lambda_2+2). $$ 

We make the observation that the $su(3)$  arity of any given irreducible
module can be
determined through 
\bea ar(\Lambda)&=& (\Lambda_1-\Lambda_2)({\rm mod}\,3) \no \\
&=& (\Lambda_1+2\Lambda_2)({\rm mod}\,3) \no \\  
&=& (-2\Lambda_1-\Lambda_2)({\rm mod}\,3)  \eea \ 
We immediately find that 
\beq ar(\Lambda)=\chi_{\Lambda}(C_3)({\rm mod}\,3). 
\label{su3arity} \eeq 

Set $\Upsilon=\exp(2\pi i/3)$. For any $V(\Lambda)$ we have 
$$\pi_{V(\Lambda)}\l(\Upsilon^{C_3}\r)=\Upsilon^{ar(\Lambda)}\cdot\I. $$ 
We now seek to find a Casimir operator which allows us to identify the
$su(3)$ singlet (with arity zero) as bosonic (parity zero), 
while the defining module (arity one) and its
dual (arity two) are identified as fermionic (parity one). We suppose that all
modules of zero arity are bosonic, and fermionic otherwise. 
A function which maps a highest weight of a module 
to its parity $p(\Lambda)$ is  
$$p(\Lambda)=\frac{4}{3}\sin^2\l( \frac{\pi.ar(\Lambda)}{3}\r).$$ 
In view of (\ref{su3arity}) we now set 
\beq K=\frac{4}{3}\sin^2\l(\frac{\pi.C_3}{3}\r) \label{su3K} \eeq  
which satisfies $\e(K)=0,\,S(K)=K$ and  
$$\chi_K(\Lambda)=p(\Lambda). $$ 

Defining
\bea
l(x)&=&\sin\l(\frac{2\pi x}{3}\r) \no \eea 
the following relation holds   
\bea
p(x+y)&=&p(x)+p(y)-\frac{3}{2}p(x)p(y)+\frac{2}{3}l(x)l(y). \no \eea 

We now find the component isomorphisms are given by  
\bea 
\tR_{x,y}&=&(-1)^{p(x)p(y)} \no \\
a_{x,y,z}&=&(-1)^{p(x)(p(y)+p(z)-p(y+z))}\no \\
&=&(-1)^{p(x)k(y,z)} \no \\
q_{w,x,y,z}&=&(-1)^{k(w,x)k(y,z)} \no \eea 
where 
\beq k(x,y)=\frac{3}{2}p(x)p(y)-\frac{2}{3}l(x)l(y) \label{k} \eeq  
Evaluating these yields that all component isomorphisms are the identity
except for 
\bea 
&&\tR_{1,1}~~~~~~a_{1,1,1}~~~~~~q_{1,1,1,1} \no \\
&&\tR_{1,2}~~~~~~a_{1,2,2}~~~~~~q_{1,1,2,2} \no \\
&&\tR_{2,1}~~~~~~a_{2,1,1}~~~~~~q_{2,2,1,1} \no \\
&&\tR_{2,2}~~~~~~a_{2,2,2}~~~~~~q_{2,2,2,2} \no \eea 
each of which is the negative of the identity. This result is in complete
agreement with \cite{joyce-s}. We identify 
${\overline{\mathbf{mod}}}^V_K(U(sl(3)))$ as the subcategory of modules
with zero arity. 

Consider a fundamental fermionic particle whose states are given by the
defining module $V=V(1,0)$ of $su(3)$ with no other degrees of freedom.    
Using the results above, any observable state is one which is an element
of a $U(su(3))$-module with zero arity. The simplest such example is
obtained by taking the three-fold tensor product of defining modules 
which admits the decomposition 
\beq ((V(1,0)\tp V(1,0)) \tp V(1,0)) 
= V(3,0)\oplus 2\times V(1,1) \oplus V(0,0). \label{su3decomp} \eeq  
Of the above modules, the states in $V(3,0)$ are completely symmetric,
the singlet $V(0,0)$ is completely antisymmetric while $V(1,1)$ is
mixed. As the statistic is fermionic, exclusion requires that the 
observable states are antisymmetric. Thus the only observable state arising
in (\ref{su3decomp}) is
the singlet. Exclusion also prohibits any observable states being
obtained from higher orders of tensor products. Hence the only
observable states are the colour vacuum and the colour singlet 
\beq \l|\Psi_C\r>=\sum_{i,j,k} \varepsilon_{ijk}\l(\l|i\r> \tp \l|j\r>
\r) \tp \l|k\r> \label{singlet} \eeq 
where $i,j,k$ take the values red, green or blue. Note that this state
is bosonic, even though it is a tensor product of three fermionic states. 

In order to consider a many-body system with many-degrees of freedom we
need to introduce additional quantum state labels.   
For the present case it is natural to first consider quark spin and flavour 
degrees of freedom which mediate electroweak interactions. Below we will
restrict to the case of the light quarks, for which an $su(3)$
flavour symmetry is appropriate. 

\section{The $su(2)\oplus su(3)$ case.} 
 
Our final example comes from a combination of the approaches of the previous 
two sections. We take the states of the fundamental particle to be given by
the tensor product of the 2-dimensional $su(2)$ defining module with the
$su(3)$ defining module. This gives a 6-dimensional space, denoted 
$V=V(1|1,0)$ with two
degrees of freedom, which we will refer to as spin and flavour. 
We denote the spin degrees of freedom by $+$ and $-$, and the flavour
labels are $u$ (up), $d$ (down) and $s$ (strange). A set of basis states
is given by 
\bea 
&&\l|+u\r>=\l|+\r>\tp \l|u\r>,~~~~\l|+d\r>=\l|+\r>\tp \l|d\r>,~~~~
\l|+s\r>=\l|+\r>\tp \l|s\r> \no \\
&&\l|-u\r>=\l|-\r>\tp \l|u\r>,~~~~\l|-d\r>=\l|-\r>\tp \l|d\r>,~~~~
\l|-s\r>=\l|-\r>\tp \l|s\r>,  \no \eea 
with the convention that $\l|+\r>,\,\l|u\r>$ are the highest weight
states and $\l|-\r>,\,\l|s\r>$ are the lowest weight states for $su(2)$
and $su(3)$ respectively. 

For the Casimir invariant $K$ which prescribes the statistic of the
fundamental particle we take the sum of the second order Casimir invariant
of $su(2)$ with the $su(3)$ Casimir invariant given by (\ref{su3K}). 
This has eigenvalue 
$$\chi_K(1|1,0)=3+1=4 $$
as given by (\ref{eig}) which shows that the fundamental particle is
{\em bosonic}. Next we need to identify   
${\overline{\mathbf{mod}}}^V_K(U(su(2)\oplus su(3)))$ for the present
case. As in the previous examples, we can label isomorphisms by arity indices 
which define equivalence classes. In the present case each class has two 
indices, one for the $su(2)$ arity and one for $su(3)$ arity. This leads to the 
following formulae for the component isomorphisms 
\bea 
\tR_{(x_1 x_2),(y_1 y_2)}&=&(-1)^{(x_1+p(x_2))(y_1+p(y_2))} \no \\
a_{(x_1 x_2),(y_1 y_2),(z_1 z_2)}&=&(-1)^{(x_1+p(x_2))
k(y_2,z_2)}\no \\ 
q_{(w_1 w_2),(x_1 x_2), (y_1 y_2), (z_1 z_2)} 
&=&(-1)^{k(w_2,x_2)k(y_2,z_2)} \label{q} \eea
where $k(w,x)$ is given by (\ref{k}).
Evaluating these formulae gives all the component isomorphisms as
the identity except the following which are the negative of the
identity;
\bea
&& \tR_{(0 b),(0 c)}~~~~~~a_{(0 b),(y_1 c), (z_1 c)}~~~~~~
q_{(w_1 b),(x_1 b),(y_1 c), (z_1 c)} \no\\
&& \tR_{(0 b),(1 0)}~~~~~~a_{(1 0),(y_1 b),(z_1 b)} \no\\
&& \tR_{(1 0), (0 b)} \no\\
&& \tR_{(1 0),(1 0)} \no
\eea
where the indices $b,c=1,2$ and $w_1,x_1,y_1,z_1=0,1$ in each component. 
Clearly this statistic indicates that the combined state is a
fermion only if the $su(2)$ state is a fermion and the $su(3)$
state a boson or vice versa.
Since the above expression (\ref{q}) is  independent of the $su(2)$ arity
labels, we conclude that 
${\overline{\mathbf{mod}}}^V_K(U(su(2)\oplus su(3)))$
consists of modules composed of an $su(2)$-module of either arity tensored 
with an $su(3)$-module of arity zero.  
As in the $su(3)$ case, we thus need to take a minimum three-fold tensor
product to construct observable states. We find that the decomposition
is given by 
\bea 
&&(V(1|1,0)\tp V(1|1,0))\tp V(1|1,0) \no \\ 
&&~~~~~~~=V(3|3,0)\oplus 2\times V(3|1,1)\oplus V(3|0,0)\oplus 
2\times V(1|3,0)\no \\
&&~~~~~~~~~~~~\oplus 4\times V(1|1,1)\oplus 2\times V(1|0,0).\no\eea 
Because the fundamental module is given as bosonic, only the completely 
symmetric modules above are not excluded. These symmetric 
modules are $V(3|3,0)$ 
and one of the copies of $V(1|1,1)$. The highest weight vector for
$V(3|3,0)$ is found to be 
$$\l|\Psi_{SF}\r>= \l(\l|+u\r>\tp \l|+u\r>\r) \tp \l|+u\r>. $$ 
We see that this state is the spin-flavour wavefunction for the  
$\Delta^{++}$ resonance in the spin 3/2
state \cite{fs}
. For the module $V(1|1,1)$ the highest weight state is given by 
\bea \l|\Psi_{SF}\r>&=& 2\l(\l|+u\r>\tp \l|+u\r>\r) \tp \l|-d\r>+ 
2\l(\l|+u\r>\tp \l|-d\r>\r) \tp \l|+u\r>+2 \l(\l|-d\r>\tp \l|+u\r>\r)
\tp \l|+u\r> \no \\
&&~~~~-\l(\l|+u\r>\tp \l|+d\r>\r) \tp \l|-u\r> 
-\l(\l|+u\r>\tp \l|-u\r>\r) \tp \l|+d\r>- \l(\l|-u\r>\tp \l|+d\r>\r)
\tp \l|+u\r> \no \\
&&~~~~-\l(\l|+d\r>\tp \l|+u\r>\r) \tp \l|-u\r> 
-\l(\l|+d\r>\tp \l|-u\r>\r) \tp \l|+u\r>- \l(\l|-u\r>\tp \l|+u\r>\r)
\tp \l|+d\r>.  \no \eea 
This state is the spin-flavour wavefunction for the proton in the spin
1/2 state \cite{huang}. 
In both cases above the colour-spin-flavour wavefunction is given by 
\cite{fs,huang} 
\beq  \l|\Psi_{CSF}\r>=\l|\Psi_{C}\r>\tp \l|\Psi_{SF}\r>. \label{csf} 
\eeq 

Using the dimension formula
$${\rm dim}\l(V(\Lambda_1|\Lambda_2,\Lambda_3)\r)
=\frac{1}{2}(\Lambda_1+1)(\Lambda_2+1)(\Lambda_3+1)(\Lambda_2
+\Lambda_3+2)$$ leads to 
\bea 
{\rm dim}\l(V(3|3,0)\r)&=&40 \no \\
{\rm dim}\l(V(1|1,1)\r)&=&16. \no \eea 
We can immediately recognise $V(1|1,1)$ as the baryon spin-1/2 $N$-octet while 
$V(3|3,0)$ is the baryon spin-3/2 $\Delta$-decuplet \cite{gn,fs,huang}. 
Note that both these modules are {\em fermionic}, as needed,  
which is determined by the eigenvalues of $K$ being odd (3 for the octet and 
15 for the decuplet). This is somewhat counter intuitive as the
fundamental
particles are bosonic. This example again illustrates how our definitions of 
the terms boson and fermion are essentially different from the usual 
interpretation, which is a consequence of the inherent non-associative
structure. Since the colour singlet (\ref{singlet}) is bosonic, 
the  
colour-spin-flavour states of the form (\ref{csf}) are all fermionic for
both the octet states and decuplet states. 

By our criteria, both the octet and the decuplet are elementary. In contrast 
to our previous examples, taking higher order tensor products yields many
more observable states, however none of them are elementary. In each case 
the number of fundamental modules in the tensor product must be a multiple of 
three to ensure that the $su(3)$ modules have zero arity. It is interesting 
to note that at this level the category of $(su(2)\oplus su(3))$-modules 
{\em is} associative. This result is entirely
consistent with the Eightfold Way of Gell-Mann and Ne'eman \cite{gn}, 
which in essence 
decrees that all hadronic states are contained in multiplets which arise 
through repeated tensor products of the 8-dimensional adjoint module
of $su(3)$. Since the adjoint module has arity zero, then so do all other 
modules in the Eightfold Way.

\section{Conclusion} 

We have introduced the operation of twining which, when applied to the 
universal enveloping algebras of semi-simple Lie algebras $U(g)$, gives rise
to symmetric, pre-monoidal categories of $U(g)$-modules, 
as defined in \cite{joyce-q}. 
By associating the state space of a fundamental particle with an 
irreducible $U(g)$-module we formulated a principle of exclusion and 
confinement for ensembles of indistinguishable particles 
as an invariance with respect to the action of the symmetric group. 
Through this procedure
we recover the spin statistics theorem for ensembles with
spin degrees of freedom.  
For quarks with $su(3)$ colour, $su(2)$ spin and $su(3)$ 
flavour degrees of freedom, our analysis identifies that each elementary
state  
is a {\em m\'enage \`a trois} of quarks which is a member of the spin-1/2 
$N$-octet or the spin-3/2 $\D$-decuplet. 
Extending this approach to obtain the observed meson multiplets remains 
as future work, for in this instance one needs to consider both fundamental
particles and antiparticles. A step in this direction is given in
\cite{ijl}. 

Our results support the consensus that quarks are confined.  
In our formulation this 
is a consequence of symmetry and statistics, not a dynamical effect 
due to the presence of a
confining force.

\begin{flushleft}
\bf{Acknowledgements} 
\end{flushleft}
PSI is a Postdoctoral Fellow supported by the Japanese Society for the
Promotion of Science.
WPJ acknowledges the support of the New Zealand Foundation for Research, 
Science and Technology (contract number UOCX0102).
JL thanks the Graduate School of Mathematical
Sciences, The University of Tokyo and the Department of Physics and
Astronomy, The University of Canterbury for generous hospitality and 
acknowledges financial support from the Australian Research
Council.

\end{document}